# Les déterminants macroéconomiques de la défaillance des emprunteurs : Cas d'une banque marocaine

# The macroeconomics determinants of default of the borrowers: The case of Moroccan bank

**Anas YASSINE, Doctorant en Sciences de Gestion, Laboratoire Management, Systèmes Financiers et Gestion des Risques (MSFGR) Université Hassan II - Casablanca- Maroc yassine.anas1@gmail.com**

**Abdelmadjid IBENRISSOUL, Professeur de l'enseignement supérieur, Laboratoire Ingénierie Scientifique des Organisations (ISO) Université Hassan II – Casablanca- Maroc a - ibenrissoul@hotmail.fr**

**Résumé**

Cet article se propose d'explorer une approche empirique pour détecter les déterminants macroéconomiques de la défaillance des emprunteurs. Ceci, dans l'objectif de mesurer et expliquer l'impact des évolutions défavorables de la conjoncture économique sur la dégradation de la qualité du portefeuille de crédit.

Dans notre étude nous avons jeté la lumière sur la problématique de l'aggravation des taux de défaut, en s'inspirant des travaux précurseurs sur la défaillance des emprunteurs. Nos résultats démontrent que les récessions économiques sont au cœur de la problématique de la solvabilité des emprunteurs. De ce fait, les banques doivent surveiller d'une façon active les évolutions des facteurs de risques macroéconomiques, et redéfinir leurs dispositifs de gestion des risques traditionnellement inscrits dans le cadre des mesures micro prudentielles.

**Abstract**

This article aims to explore an empirical approach to analyze the macroeconomics determinants of default of borrowers. For this purpose, we have measured the impact of the adverse economic conditions on the degradation of the credit portfolio quality.

In our paper, we have shed more light on the question of the aggravation of default rate. For this, we have undertaken econometric modeling of the default rate distribution of a Moroccan bank while we inspired from some studies carried out. Our findings demonstrate that the decline in the economic situation has a positive impact on default of borrowers. Hence, the bank also has responsibility for monitoring the adverse economic conditions.

## Introduction

La réglementation bancaire marocaine a connu des évolutions sensibles, qui ont impacté les banques de façon systématique. Le secteur bancaire marocain ne présentant aucune particularité quant à l'assimilation de ces mutations qui s'inscrivent dans un souci de rénovation et d'adaptation à un environnement et à des conditions de fonctionnement connaissant de profondes mutations. La crise des subprimes (2008) nous a bien rappelé l'importance d'une supervision efficace du système bancaire par les autorités de supervision et de régulation. D'autant plus que l'année 2008 a été marquée, sur le plan international, par l'amplification de cette crise financière qui est née de l'effondrement du marché américain des prêts hypothécaires à risque. Ses effets se sont rapidement propagés à l'économie réelle à l'échelle mondiale. Plusieurs pays développés sont entrés en récession, alors que le rythme de croissance des pays émergents a sensiblement ralenti.

Dans ce contexte, La faillite de certaines institutions financières d'envergure systémique a ébranlé la confiance des opérateurs des marchés financiers, contribuant, par là même, à la montée de l'aversion au risque et à la très nette perturbation des mécanismes des marchés de crédit. Face à ces évolutions, les banques centrales et les autorités de supervision financière se sont activées pour circonscrire les incidences de ces turbulences et en limiter la propagation en s'appuyant sur le risque de crédit qui compte parmi les plus anciens et les plus graves risques inhérents au cœur des métiers bancaires.

Aujourd'hui on peut se permettre de dire que les banques marocaines sont bien contrôlées et bien gérées, tout danger de faillites étant écarté grâce à une réglementation soigneusement élaborée et au déploiement des dispositifs de gestion des risques financiers. Face aux restrictions réglementaires rigoureuses auxquelles les banques se trouvent confrontées, dans quelle mesure peut-on-dire que la question du risque de défaut de paiement peut-elle être réglée ?

On ne doit pas se contenter de constater l'efficacité des dispositifs de gestion des risques mis en place sans chercher à déterminer les facteurs d'influence externes pouvant impacter la solvabilité des clients de la banque. En effet, La question de la dégradation de la conjoncture économique est au cœur de la problématique de la défaillance des emprunteurs.

Dans ce contexte, le présent article tente d'examiner dans quelle mesure les évolutions de l'environnement macroéconomique peuvent accroitre le taux de défaut des clients de la banque. Ce travail se décline en trois parties. La première jette la lumière sur les études empiriques des déterminants macroéconomiques de l'insolvabilité des emprunteurs. La deuxième donne un aperçu des différentes sources et caractéristiques de donnée utilisées dans l'étude, ainsi que la méthodologie employée pour la réalisation de cette dernière. La troisième partie du travail expose les résultats obtenus et les analyses qui en découlent.

## 1. Etudes empiriques sur les facteurs macroéconomiques explicatifs de la défaillance des emprunteurs

L'examen de la littérature empirique s'intéressant aux défauts de paiement, montre que la majorité des études réalisées en vue d'appréhender la dimension de la défaillance des emprunteurs se concentrent sur des aspects microéconomiques, en passant par des analyses de données cross-sectionnelles (en coupe transversale) sur les individus Fox (2010). La majorité des études réalisées sont soit des études américaines soit des études canadiennes. De plus, Les méthodologies déployées dans le cadre de ces études sont différentes. Les objectifs ultimes de notre investigation de ce courant de recherche est d'analyser les principaux résultats donnés par tout un chacun, et justifier notre choix en matière de la sélection des facteurs macroéconomique susceptibles d'expliquer l'insolvabilité des clients de la banque.

### 1.1. Les indicateurs macroéconomiques explicatifs de l'évolution du nombre de défaut aux Etats-Unis

L'ancienne crise des subprimes (2008) aux Etats Unis nous a donné une idée sur les déterminants de la défaillance des emprunteurs, elle a contribué au développement et à l'enrichissement de la littérature empirique par la confirmation ou l'infirmation de certains résultats antérieurs. Dans la même voie, plusieurs auteurs ont tenté de déterminer les facteurs explicatifs des faillites, Yeager (1974) est considéré comme l'un des premiers à avoir réalisé une étude empirique cherchant à déterminer les facteurs macroéconomiques expliquant les faillites des individus aux Etats-Unis. Il a conçu un modèle de croissance du nombre de faillite sur la base des données concernant les ménages américains pour une période allant de 1950 à 1970. Il a parvenu à expliquer 94% de la croissance par une seule variable explicative : le ratio d'endettement des ménages retardé de 6 mois (c'est à dire le rapport entre les crédits à la consommation et le revenu disponible). Il a essayé d'augmenter la capacité prédictive de son modèle en incluant le taux de chômage, retardé de plusieurs périodes mais l'inclusion de cette variable a été peu concluante. Sullivan (1983) a tenté de mettre à jour l'étude faite par Yeager(1974) en élargissant la période d'analyse de 1950 à 1982. Son étude a confirmé le rapport entre le ratio d'endettement et le taux de faillite. Il a également montré qu'une autre variable expliquait le taux de défaut à savoir, le Consumer Sentiment Index. Ce dernier mesure indirectement la volonté des ménages de rembourser leurs dettes (Mathilde FOX, 2010, p. 92).

Agarwalet Liu (2003) ont réalisé une étude en se focalisant essentiellement sur la relation entre le taux de chômage et la faillite des consommateurs. Leur étude confirme l'existence d'une relation positive entre le taux de chômage et l'insolvabilité des consommateurs.

Paquin P. et Weiss M.S. (1998) ont suivi une démarche un peu différente afin de modéliser le taux de défaut des consommateurs. Ils ont parvenu à l'expliquer à hauteur de 98% à l'aide d'un modèle autorégressif (AR) incluant les variables suivantes : l'offre de crédit à la consommation, la capacité de remboursement mesurée par le ratio des dettes sur le revenu, les conditions du marché de l'emploi et le niveau des taux d'intérêt.

### 1.2. Les indicateurs macroéconomiques explicatifs de l'évolution du nombre de défaut au Canada

La problématique de la défaillance des emprunteurs est une question multidimensionnelle qui n'a pas de frontière. En effet, parallèlement aux études américaines sur la défaillance des emprunteurs, plusieurs études ont été réalisées dans le contexte canadien en vue de tenter d'expliquer l'impact des évolutions défavorables de la conjoncture économique sur la solvabilité des emprunteurs.

O’Neil (1998) constate que la croissance des prêts bancaires non performants peut être expliquée en partie par l’état général de la conjoncture économique.

A l’instar des travaux d’ O’Neil (1998), Swart et Anderson (1998) ont tenté d’expliquer la croissance du nombre de faillites à l’aide de l’évolution du PIB réel et du taux de chômage. Ces facteurs ont donné des résultats significatifs (Mathilde FOX, 2010, p. 97).

Chang Soo K. et al (1998) ont suivi une approche théorique différente. Ils ont estimé un modèle de prédiction du taux de défaut en se focalisant essentiellement sur l’excédent de liquidité dont dispose l’entreprise. Le constat de leur travail postule que les entreprises qui risqueraient de devenir insolvables sont en principe celles qui épargnent moins.

En effet, La défaillance des emprunteurs aux États-Unis et au Canada est un phénomène spécifique à ces deux pays et en l’occurrence les diverses conclusions qu’on a tirées de l’examen de la littérature empirique vont nous servir uniquement de guide pour que nous puissions orienter notre choix en matière de sélection des variables susceptibles d’expliquer la dégradation de la qualité du portefeuille de crédit appartenant à une banque marocaine.

## 2. Données et méthodologie

### 2.1. Sources et caractéristiques de données

Pour vérifier l’impact des évolutions défavorables de la conjoncture économique sur l’accroissement du taux de défaut des clients de la banque, on a tenté de recenser toutes les variables disponibles et qui se rapprochent le plus de celles utilisées dans la littérature empirique. La démarche de recherche de données qu’on a poursuivie nous a permis de recueillir deux types d’informations différentes :

Des informations internes à la banque sur la composition du portefeuille de crédit et les taux de défaut[1]y afférents. Ces informations proviennent des états financiers et des bases de données contenant la notation des clients par classe d'exposition.

**Tableau 1** : Récapitulatif de la statistique descriptive des variables internes

| | Crédits à la clientèle Corporate | Crédits à la clientèle Retail | Crédits aux sociétés de financement | Dépôts de la clientèle | Total crédits | Taux de défaut |
|---|---|---|---|---|---|---|
| **Mean** | 34594.01 | 16491.13 | 6836.153 | 71668.10 | 61028.97 | 0.053832 |
| **Median** | 35503.70 | 17732.77 | 7776.414 | 75830.71 | 63566.39 | 0.050350 |
| **Maximum** | 58020.91 | 25999.30 | 10216.41 | 97184.94 | 98104.44 | 0.087700 |
| **Minimum** | 18817.92 | 5379.090 | 3460.610 | 42926.86 | 30316.21 | 0.038300 |
| **Std. Dev.** | 11376.54 | 6671.628 | 2235.582 | 16718.52 | 20571.00 | 0.015185 |
| **Observations** | 28 | 28 | 28 | 28 | 28 | 28 |

Source : composé par l'auteur

Des informations externes à la banque correspondant à des variables macroéconomiques représentatives de la sphère économique et financière marocaine. Ces données sont collectées en fréquence trimestrielle sur sept ans (de 2005 à 2011) en passant par le site du Haut-Commissariat au Plan(HCP), et celui du Bank Al Maghrib (Banque centrale).

**Tableau 2 :** Synthèse des variables externes

| Variables exogènes retenues | Signes attendus | Référence bibliographique | Lien avec le taux de défaillance des emprunteurs |
|---|---|---|---|
| LogPIB_vol : le logarithme du PIB réel en volume | Négatif | • O'neil (1998),<br>• Swart et Anderson (1998), | L'accroissement du PIB se traduit par l'augmentation des revenus des agents économiques, et donc une amélioration de la capacité de remboursement des emprunteurs. En revanche, la baisse du niveau du PIB devrait en principe faire accroitre les défauts de paiement, d'où un signe négatif. |

[1]Créances cotées 11 et 12 divisées par les créances cotées 8,9 et 10

| | | | |
|---|---|---|---|
| Tx_chom : taux de chômage | Positif | • Yeager (1974),<br>• Swart et Anderson (1998),<br>• Agarwal et Liu (2003), | Le rebond du taux de chômage accroît la probabilité que l'emprunteur se retrouve sans emploi et éprouve des difficultés pour rembourser son prêt. Le signe attendu de la relation entre le chômage et les taux de défaut est positif. |
| Tx_débi : taux d'intérêt réel | Positif | • Paquin P. et Weiss M.S. (1998), | Une augmentation du taux d'intérêt débiteur pourrait engendrer une fragilisation de la situation financière des emprunteurs qui auront du mal à couvrir leurs emprunts, d'où un signe positif. |
| Eparg_Vol : Epargne nationale brute en volume | Négatif | • Chang soo K. et al (1998), | Pour cette variable, le raisonnement concerne le surplus de la consommation par rapport au revenu. Ce surplus traduirait une disponibilité des fonds qui, quand elle augmente impacte positivement la solvabilité des emprunteurs. D'où un signe négatif. |
| USD & EUR/MAD : Taux de change du dirham marocain par rapport au dollar américain et à l'euro | Ambigu | • Khemraj et Pacha (2009),<br>• Drehman M. (2005), | Une variation positive du taux de change par rapport au dollar américain ou l'euro, signifie que le dirham Marocain baisse. Alors, on peut interpréter cette baisse comme étant un signe de faiblesse relative de l'économie marocaine. Cependant, on sait également que si le dirham marocain se déprécie les exportations en bénéficient, ce qui devrait à son tour améliorer l'économie marocaine. Donc on ne peut pas avoir une idée claire sur le sens de la relation. |
| Tx_infla : taux d'inflation | Ambigu | • Figlewsk et al. (2006),<br>• Dionne et al. (2007), | L'impact de l'inflation sur le taux défaut est théoriquement très ambigu. En effet, la hausse d'inflation agit comme un frein à l'économie d'où un signe positif. Par contre, les débiteurs dont la dette est en terme nominal verront leur dette en terme réel diminuer avec l'augmentation de l'inflation. |

Source : élaboré par l'auteur

Le choix des variables exogènes retenues dans la modélisation du taux de défaut découle à la fois de la littérature empirique sur les déterminants de la défaillance des emprunteurs. Par ailleurs, L'examen exhaustif de données met évidence deux contraintes :

- Il n'existe pas toujours de parfait équivalent marocain aux indicateurs qui ont servi dans les études citées précédemment.
- De nombreuses données n'existent pas en fréquence trimestrielle. Or notre série de taux de défaut est trimestrielle.

### 2.2. Méthodologie

Compte tenu de la nature quantitative des variables exogènes sélectionnées comme candidates potentiellement explicatives de l'accroissement du taux de défaut (voir tableau 2), il s'avère pertinent d'opter pour une régression linéaire multiple (RLM) par la méthode des moindres carrées ordinaires (MCO). Ceci pour spécifier un modèle économétrique de prédiction de la dégradation de la qualité du portefeuille de crédit. La spécification du modèle est faite en utilisant la technique d'estimation dite pas à pas (Stepwise). Cette procédure comporte deux approches de sélection :

Une approche descendante également appelée Backward. Elle consiste à estimer les statistiques F du modèle qui inclut toutes les variables potentiellement explicatives de la variable endogène (Taux de défaut). A chaque étape du processus une variable est supprimée, il s'agit de la variable qui contribue le moins au modèle. Ces démarches sont poursuivies une à une jusqu'à ce qu'il ne reste plus que des variables significatives à un niveau de significativité[2] spécifié dans le modèle.

[2] Dans cette étude on travaille avec un niveau de confiance de 95%

Dans la sélection ascendante ou forward on renverse la vapeur. On commence par un modèle sans aucune variable, après on introduit une par une les variables à coefficient de corrélation partiel significatif. Le principal défaut de cette méthode est que quand une variable est entrée elle y reste même si elle devient par la suite non significative. D'ailleurs, c'est la raison pour laquelle on n'a pas opté pour cette démarche de sélection.

## 3. Résultats et discussion

### 3.1. Etude de la stationnarité des séries (Test Dickey Fuller augmenté)

Les observations dont on dispose pour la réalisation de cette étude constituent une série temporelle qui doit être soigneusement analysée. En effet, L'analyse d'une série temporelle tente généralement d'aborder les deux points suivants :

- La série contient-elle (ou non) une tendance ? Est-elle (ou non) stationnaire ?
- La volatilité est-elle constante (homoscédasticité) ou variable ?

Le test de Dickey-Fuller permet de mettre en évidence le caractère stationnaire ou non d'une chronique par la détermination d'une tendance déterministe (processus TS : Time Stationary) ou stochastique (processus DS : Differency Stationary).

Les modèle servant de base à la construction de ce test sont au nombre de trois. Le principe du test est simple : si l'hypothèse $H_0$:$\emptyset = 1$ est retenue dans l'un de ces trois modèles, le processus est alors non stationnaire.

- Modèle sans constante et sans tendance : $\Delta\ X_t$=$\emptyset\ -1$+$\varepsilon_t$ (1)
- Modèle avec constante sans tendance : $\Delta X_t$=$\emptyset X_t$$-1$+$c$ + $\varepsilon_t$ (2)
- Modèle avec constante et tendance : $\Delta\ X_t$=$\emptyset X_t$$-1$+$c$+$b_t$+$\varepsilon_t$ (3)

Si l'hypothèse $H_0$est vérifiée, la chronique n'est pas stationnaire quel que soit le modèle retenu. Dans le dernier modèle (3), si on accepte $H_1 : \emptyset < 1$ et si le coefficient b est significativement différent de 0, alors le processus est un processus TS ; on peut le rendre stationnaire en calculant les résidus par rapport à la tendance estimée par les moindres carrées ordinaires.

**Tableau 3** : synthèse test de stationnarité (Test ADF**)**

| Séries | Hypothèse nulle | Hypothèse alternative | Valeur de la Statistique ADF | Modèle choisi | Seuil Alpha à 5% | Niveau d'intégration |
|---|---|---|---|---|---|---|
| **LogPIB_vol** | Non stationnaire | stationnaire | -4.471762 | Avec constante et sans tendance | -3.020686 | I(0) |
| **Tx_chom** | Non stationnaire | stationnaire | -2.620002 | Sans constante et sans tendance | -1.958088 | I(0) |
| **Tx_débi** | Non stationnaire | stationnaire | -3.463997 | Avec constante et sans tendance | -2.976263 | I(0) |
| **Eparg_vol** | Non stationnaire | stationnaire | -2.815412 | Avec constante et sans tendance | -2.976263 | I(1) |
| **EUR/MAD** | Non stationnaire | stationnaire | -2.991123 | Avec constante et sans tendance | -2.976263 | I(0) |
| **USD/MAD** | Non stationnaire | stationnaire | -3.237877 | Avec constante et sans tendance | -2.981038 | I(0) |
| **Tx_infla** | Non stationnaire | stationnaire | -2.996883 | Sans constante et sans tendance | -1.954414 | I(0) |
| **Tx_Déf** | Non stationnaire | stationnaire | -2.672175 | Sans constante et sans tendance | -1.953858 | I(0) |

Source : élaboré par l'auteur

Ce tableau révèle les résultats du test de stationnarité appliqué sur les séries retenues dans la modélisation. On constate que toutes les séries sont stationnaires sauf celle qui correspond à l'épargne nationale brute. La non stationnarité de cette dernière pourrait être à l'origine d'une régression fallacieuse faisant apparaître des résultats erronés. Pour remédier à ce problème on a utilisé le test AugmentedDickey Fuller en vue de stationnariser la série en question.

**Graphique 1** : évolution de l'épargne nationale brute en fréquence trimestrielle

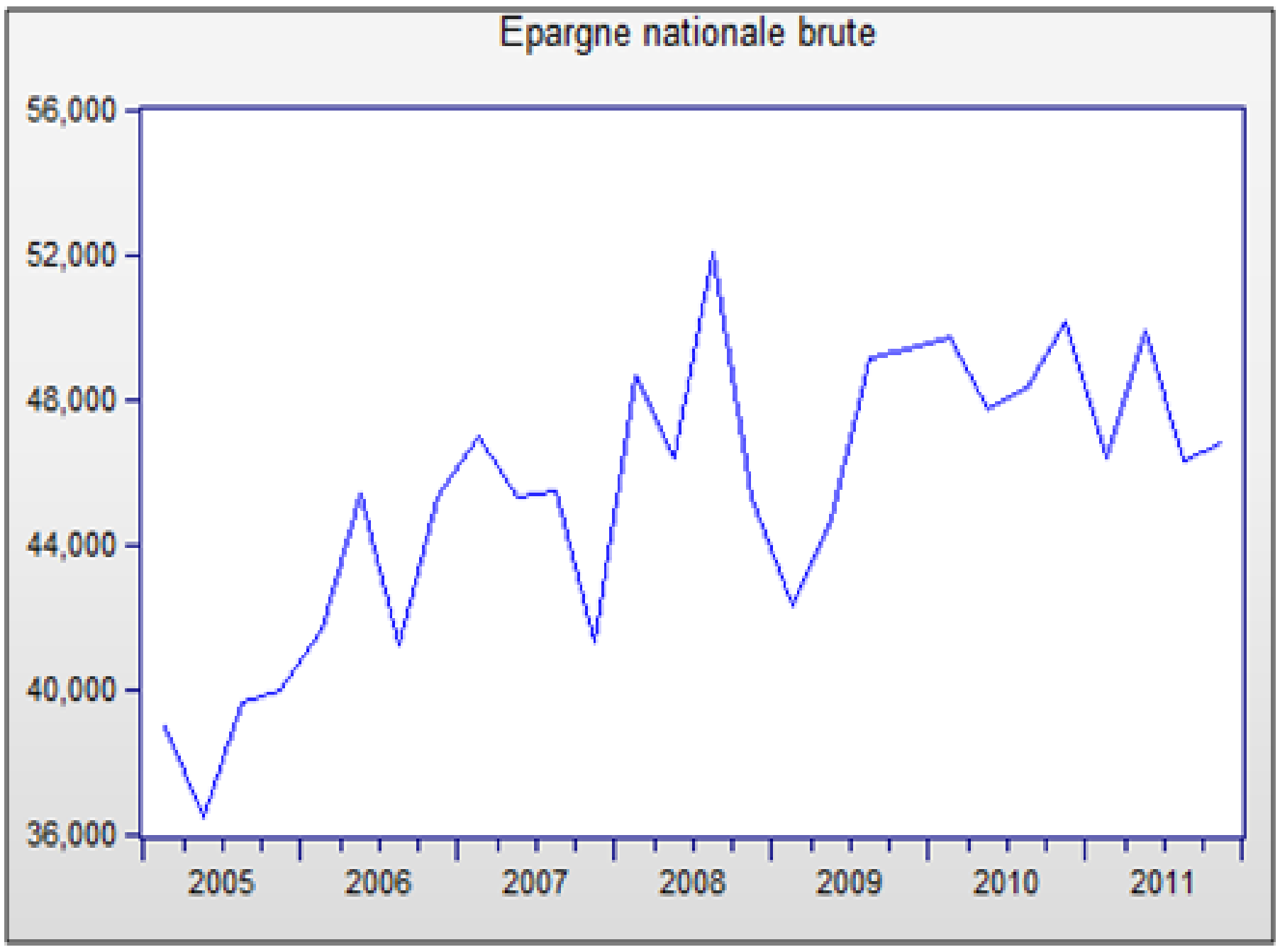

D'après le test ADF La série Eparg est un processus non-stationnaire de manière stochastique (processus DS : DifferencyStationary). La non-stationnarité est alors stochastique par le biais des perturbations dues au fait que la variance n'est pas constante et affectée par le temps.

**Tableau 4 :** test de stationnarité de la série (Eparg): modèle avec constante et tendance

| | | | t-Statistic | |
|---|---|---|---|---|
| Augmented Disky-Fuller test statistic | | | -3.111898 | |
| Test critical values : | 1% level | | -4.374307 | |
| | 5% level | | -3.603202 | |
| | 10% level | | -3.238054 | |
| Variable | Coefficient | Std. Error | t-Statistic | Prob. |
| Eparg(-1) | -1.045239 | 0.335885 | -3.111898 | 0.0055 |
| C | 43917.41 | 13502.64 | 3.252505 | 0.0040 |
| TREND(2005Q1) | 282.2573 | 145.6176 | 1.938346 | 0.0668 |

On remarque à partir du test effectué que la tendance n'est pas significative, puisqu'elle présente une erreur de 6.68% relativement supérieure au seuil tolérable de 5%. Par conséquent, le modèle approprié pour tester la stationnarité de la série (Eparg) est celui qui ne présente pas de tendance, d'où le recours à la deuxième étape.

**Tableau 5 :** test de stationnarité de la série (Eparg) : modèle avec constante et sans tendance

| | | | t-Statistic | |
|---|---|---|---|---|
| Augmented Disky-Fuller test statistic | | | -2.815412 | |
| Test critical values : | 1% level | | -3.699871 | |
| | 5% level | | -2.976263 | |
| | 10% level | | -2.62742 | |
| Variable | Coefficient | Std. Error | t-Statistic | Prob. |
| Eparg(-1) | -0.425279 | 0.151054 | -2.815412 | 0.0094 |
| C | 19563.78 | 6871.34 | 2.847171 | 0.0087 |

D'après le résultat du test réalisé, il est possible de constater que la constante est significative du fait qu'elle est égale à 0.087% qui est largement inférieur au seuil tolérable de 5%. Il en résulte que le modèle avec constante et sans tendance semble être le plus approprié pour tester la stationnarité de la série (Eparg). Cependant l'application du test AugmentedDickey Fuller (ADF) fait ressortir la présence d'une racine unitaire dans la série (Eparg), donc la série est non stationnaire et cela s'explique par la statistique ADF qui égale à (-2.815412) est supérieure à la valeur critique au seuil de 5% qui égale à (-2.976263). Pour remédier à ce problème on doit appliquer le test ADF sur la série en différence première, et ce afin de tester son ordre d'intégration.

**Tableau 6** : test de stationnarité de la série (Eparg): modèle avec constante et sans tendance « en différence première »

| | | | t-Statistic | |
|---|---|---|---|---|
| Augmented Disky-Fuller test statistic | | | -8.183802 | |
| Test critical values : | 1% level | | -2.656915 | |
| | 5% level | | -1.954414 | |
| | 10% level | | -1.609329 | |
| Variable | Coefficient | Std. Error | t-Statistic | Prob. |
| D (EPARG(-1)) | -1.446299 | 0.176727 | -8.183802 | 0.0001 |
| C | 19425.65 | 5347.264 | 2.695741 | 0.0094 |

D'après le test réalisé en constate que la série est stationnaire en différence première vu que la statistique ADF qui égale à (-8.183802) est inférieure à la valeur critique au seuil de 5% qui égale à (-1.954414). Alors, on peut se permettre de dire que la série (Eparg) est stationnaire en différence première sans tendance et avec constante.

### 3.2. Présentation des modèles spécifiés

| Variable endogène | Taux de défaut | | | | |
|---|---|---|---|---|---|
| **Spécifications** | **Spé 1** | **Spé 2** | **Spé 3** | **Spé 4** | **Spé 5** |
| **Variables exogènes** | | | | | |
| **Constante** | 0.416937 (1.513984) | 0.431579 (1.590783) | 0.466457 (1.698653) | 0.515858 (1.888793) | 0.597018 (2.778288) |
| **LogPIB_vol** | -0.074978 (-1.274256) | -0.077556 (-1.336769) | -0.089702 (-1.567039) | -0.126252 (-2.633621) | -0.140413 (-3.710337) |
| **Tx_chom** | 0.351931 (1.261184) | 0.290748 (1.107496) | 0.299931 (1.148332) | 0.095191 (0.494703) | |
| **Tx_débi** | 0.678876 (1.240547) | 0.726717 (1.354352) | 1.159415 (2.442536) | 1.379098 (3.151783) | 1.399099 (3.262983) |
| **Eparg_vol** | 3.07E-07 (0.709377) | 2.91E-07 (0.680771) | | | |
| **MAD_EUR** | -0.011930 (-1.175204) | -0.012074 (-1.204232) | -0.010899 (-1.149839) | | |
| **MAD_USD** | 0.009493 (2.176803) | 0.009843 (2.299070) | 0.008190 (2.083489) | 0.010990 (3.536005) | 0.010964 (3.585125) |
| **Tx_infla** | -0.122092 (-0.718989) | | | | |
| **R2** | 0.829580 | 0.824943 | 0.839164 | 0.829498 | 0.827684 |
| **DW** | 1.457385 | 1.464999 | 1.477574 | 1.158297 | 1.117226 |

Le tableau ci-dessus révèle les résultats des différentes régressions obtenues en passant par la méthode dite de sélection descendante (voir plus haut). La meilleure spécification obtenue est celle qui inclut simultanément le Logarithme du PIB réel en volume, le taux d'intérêt et le taux de change USD/MAD.

- Modèle théorique de base :

$$Y_i = \beta_0 + \beta_1 X_{i1} + \beta_1 X_{i1} + \dots + \beta_p X_{ip} + \varepsilon_i \qquad i = 1,\dots, n \qquad (4)$$

$$Txdéf = \beta_0 + \beta_1 LogPIB_Vol_1 + \beta_2 Tx_Chom_2 + \beta_3 Tx_débi_3 + \beta_4 Eparg_vol_4 + \beta_5 EUR/MAD_5 + \beta_6 USD/MAD_6 + \beta_7 Tx_infla_7 + \varepsilon_i \qquad (5)$$

- Modèle retenu :

$$Txdéf = \beta_0 + \beta_1 LogPIB_Vol_1 + \beta_2 Tx_débi_2 + \beta_3 USD/MAD_3 + \varepsilon_i \qquad (6)$$

Où : Txdéf est la variable endogène
$LogPIB_Vol_1$, $Tx_débi_2$, USD/MAD sont des variables exogènes
$\beta_1$, $\beta_2$, $\beta_3$ correspondent aux coefficients de régression
$\beta_0$ est la constante du modèle
et $\varepsilon_i$ est le terme d'erreur

Txdéf = 0.597018 -0.140413Log PIB_Vol + 1.399099 Tx_débi + 0.010964 USD/MAD (7)

Par ailleurs, on sait très bien que le R2 (R-squared) ou le coefficient de détermination mesure la qualité de l'ajustement des estimations de l'équation de régression. Dans le cas de notre modèle, la valeur de R2 est égale à 0,827684. En ramenant cette dernière en pourcentage, il est possible de l'interpréter comme suit : 82,77 % de la variabilité du Taux de défaut afférent au portefeuille étudié est expliquée par la liaison avec le logarithme du PIB réel en volume, le taux d'intérêt et le cours de change USD/MAD. Les 17,24% restants représentent les erreurs de mesures et toutes les imprécisions engendrées lors de l'expérience.

### 3.2.1. Tests de validation

#### 3.2.1.1. Test d'autocorrélation des résidus

La réalisation du test d'autocorrélation des résidus est une étape incontournable dans le processus de validation du modèle économétrique spécifié. En effet, Il existe plusieurs méthodes pour tester l'absence d'autocorrélation des erreurs dont la plus utilisée est la statistique de Durbin Watson. Cette statistique permet de mesurer l'absence de corrélation entre les résidus. En d'autre terme, elle permet de tester si deux écarts successifs se ressemblent, sa valeur est comprise entre 0 et 4. Une valeur proche de zéro signifie qu'il y a une autocorrélation positive, les valeurs situées autour de 2 montrent une absence d'autocorrélation, ainsi plus la valeur s'approche de 4, plus il y aura alors une autocorrélation négative. Cette statistique pose donc problème lorsqu'elle est proche de zéro ou de quatre.

Dans le cas de notre modèle la statistique DW = 1.117226. Alors, on peut se permettre de dire qu'il y a une absence d'autocorrélation des erreurs.

### 3.2.1.2.Test d'hétéroscédasticité

Le tester l'homoscédasticité porte sur l'analyse de la variance des erreurs de la régression.

$H_0$ : Homoscédasticité

$H_1$ : Hétéroscédasticité

**Encadré 1** : résultat du test de WHITE

Heteroskedasticity Test: White

| | | | |
|---|---|---|---|
| F-statistic | 0.309992 | Prob. F(8,19) | 0.9527 |
| Obs*R-squared | 3.232701 | Prob. Chi-Square(8) | 0.9189 |
| Scaled explained SS | 3.570239 | Prob. Chi-Square(8) | 0.8937 |

Test Equation:
Dependent Variable: RESID^2
Method: Least Squares
Date: 06/14/12 Time: 14:59
Sample: 2005Q1 2011Q4
Included observations: 28
Collinear test regressors dropped from specification

| Variable | Coefficient | Std. Error | t-Statistic | Prob. |
|---|---|---|---|---|
| C | -0.062781 | 0.080378 | -0.781070 | 0.4444 |
| TX_DEBI | 0.142037 | 1.380989 | 0.102852 | 0.9192 |
| TX_DEBI^2 | 0.561782 | 2.214563 | 0.253676 | 0.8025 |
| TX_DEBI*MAD_USD | -0.014070 | 0.011208 | -1.255418 | 0.2245 |
| TX_DEBI*LOGPIB_VOL | -0.019951 | 0.223736 | -0.089172 | 0.9299 |
| MAD_USD | 0.007133 | 0.007655 | 0.931817 | 0.3631 |
| MAD_USD^2 | -1.06E-05 | 6.16E-05 | -0.172145 | 0.8651 |
| MAD_USD*LOGPIB_VOL | -0.001171 | 0.001345 | -0.870464 | 0.3949 |
| LOGPIB_VOL | 0.011080 | 0.014284 | 0.775701 | 0.4475 |

| | | | |
|---|---|---|---|
| R-squared | 0.115454 | Mean dependent var | 3.83E-05 |
| Adjusted R-squared | -0.256987 | S.D. dependent var | 6.77E-05 |
| S.E. of regression | 7.59E-05 | Akaike info criterion | -15.88053 |
| Sum squared resid | 1.09E-07 | Schwarz criterion | -15.45232 |
| Log likelihood | 231.3274 | Hannan-Quinn criter. | -15.74962 |
| F-statistic | 0.309992 | Durbin-Watson stat | 1.373974 |
| Prob(F-statistic) | 0.952747 | | |

Source : réalisé par nous même

D'après le résultat du test, il est possible de constater que la probabilité associée au test Obs*R-squared (0.9189) est supérieure à 5%. De ce fait, on accepte l'hypothèse nulle et on rejette l'hypothèse alternative.

### 3.2.1.3.Test de normalité des résidus (Test de Jarque-Bera)

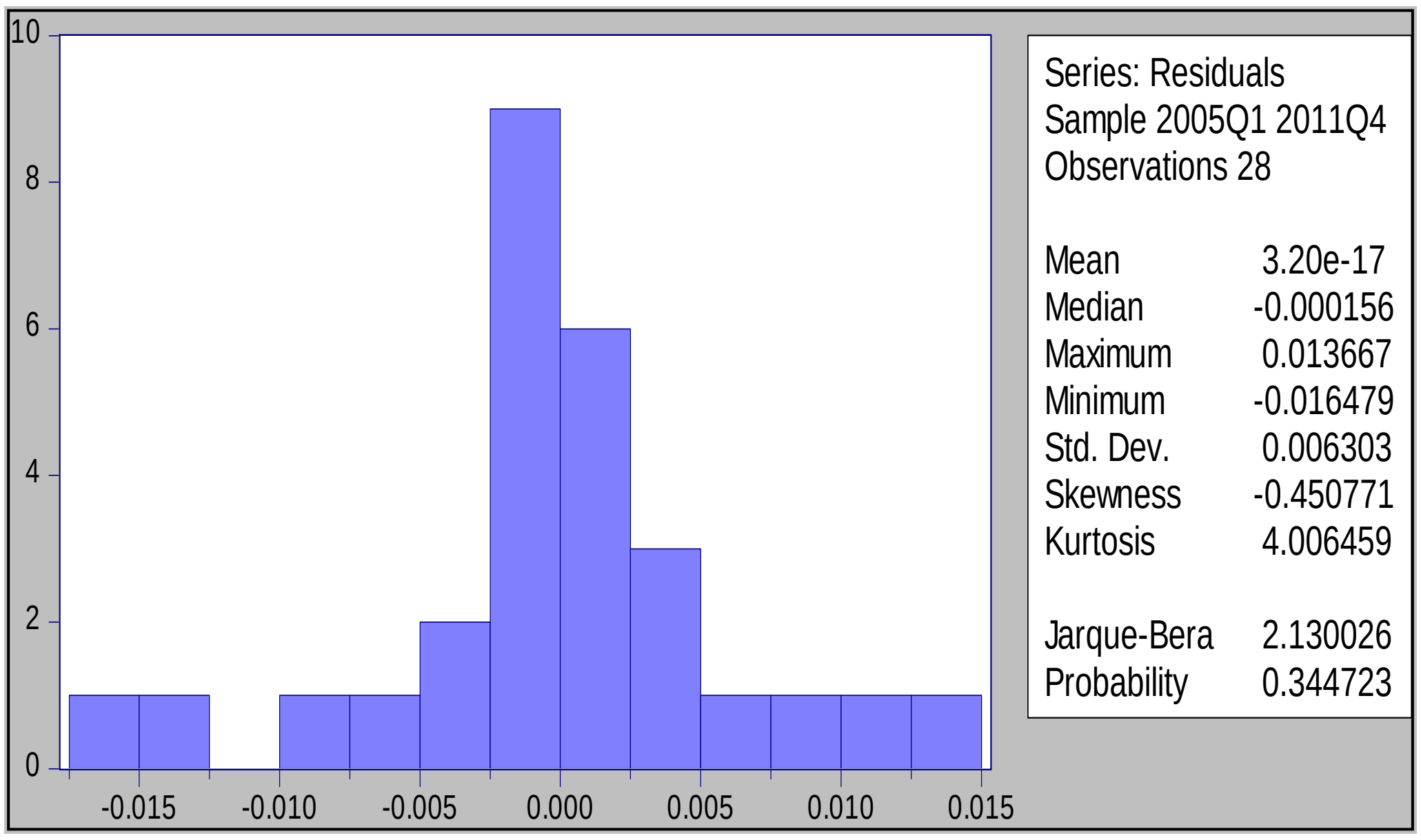

L'objectif de la réalisation du test de normalité de Jarque-Bera est de s'assurer que les résidus de notre modèle économétrique suivent une distribution normale.

Comme chaque test d'hypothèse, il faut poser une hypothèse nulle à valider :

$H_0$ : Les erreurs du modèle suivent une distribution normale.

$H_1$: Les erreurs du modèle ne suivent pas une distribution normale.

Cependant, Le test de Jarque-Bera ne teste pas à proprement parler si les données suivent une loi normale, mais plutôt si le coefficient d'aplatissement (kurtosis ) et le coefficient d'asymétrie (Skewness) des erreurs du modèle sont les mêmes que ceux d'une loi normale de même espérance et variance.

On pose donc :

$$H_0 : S = 0 \text{ et } K = 3 \qquad\qquad H_1 : S \neq 0 \text{ et } K \neq 3$$

On remarque à partir des résultats obtenus que la probabilité associée au test de Jarque-Bera (0.344723) est supérieure à 5%. Alors, on accepte l'hypothèse $H_0$ et on rejette l'hypothèse alternative.

### 3.3. Discussion

Au niveau de ce travail, on a essayé de déterminer les facteurs de risques macroéconomiques qui peuvent impacter la solvabilité des emprunteurs et accroitre leurs taux de défaut. Les résultats obtenus montrent que le PIB impacte négativement le taux de défaut des clients de la banque. Cela s'expliquer sans aucun doute par le fait que pendant les périodes d'expansion économique, les revenus des ménages et les profits des entreprises s'améliorent et, par voie de conséquence leur permettent de couvrir leurs dettes. Ce résultat est confirmé par les travaux de Swart et Anderson (1998) et O'Neil (1998). En ce qui concerne le taux d'intérêt réel, Le modèle économétrique qu'on a spécifié révèle un coefficient positif et significativement différent de zéro ($\beta_2$ = 1.39099). Cela signifie qu'une augmentation brutale des taux d'intérêts pourrait fragiliser la situation financière des emprunteurs dont les crédits sont contractés à taux variable. L'étude qui a été réalisée par Paquin P. et Weiss M.S (1998) a démontré que la hausse du taux d'intérêt impact positivement le taux de défaut des consommateurs américains. En outre, Le coefficient de la variable taux de change (USD/MAD) qui a été retenue dans notre modèle prédictif du taux de défaut est statistiquement significatif ($\beta_3$ = 0.010964), et évoque une relation positive avec la variable endogène (voir modèle retenu). Cela dit, qu'une dépréciation du dirham marocain par rapport au dollar américain peut alourdir la balance commerciale du pays, car les prix des biens et services importés vont s'envoler naturellement en monnaie locale ce qui n'est pas sans provoquer une baisse du pouvoir d'achat du consommateur marocain.

## Conclusion

Dans le présent article, nous avons examiné dans quelle mesure le risque de défaut peut être expliqué par les évolutions défavorables de l'environnement macroéconomique. Nos résultats démontrent que la croissance du PIB a un impact négatif sur le taux de défaut des clients de la banque, tandis que le taux d'intérêt réel et le taux de change l'impactent positivement. Il est maintenant clair que la dégradation de conjoncture économique est au cœur de la problématique de la défaillance des emprunteurs. De ce fait, les banques doivent surveiller d'une façon active les évolutions des facteurs de risques macroéconomiques en intégrant les stress tests dans leurs dispositifs de gestion des risques afin de s'assurer qu'elles resteraient résilientes en cas de choc macroéconomique sortant de l'ordinaire. Par ailleurs, cette analyse s'avère également pertinente pour la politique macro prudentielle, qui peut limiter le risque de crédit en agissant sur plusieurs de ces variables.

## Bibliographie